\documentclass[aps,prl,twocolumn,showpacs,floatfix,superscriptaddress]{revtex4}
\usepackage[dvips]{graphicx}
\begin{document}
\flushbottom
\title{Pressure dependence of superconducting critical temperature and upper critical field of 2H-NbS$_2$}
\author{V. G. Tissen}
\affiliation{Institute of Solid State Physics, Chernogolovka, 142432 Moscow Region, Russia}
\affiliation{Laboratorio de Bajas Temperaturas, Departamento de F\'isica de la Materia Condensada, Instituto de Ciencia de
Materiales Nicol\'as Cabrera, Facultad de Ciencias \\ Universidad Aut\'onoma de Madrid, E-28049 Madrid, Spain}

\author{M. R. Osorio}
\affiliation{Laboratorio de Bajas Temperaturas, Departamento de F\'isica de la Materia Condensada, Instituto de Ciencia de
Materiales Nicol\'as Cabrera, Facultad de Ciencias \\ Universidad Aut\'onoma de Madrid, E-28049 Madrid, Spain}

\author{J. P. Brison}
\affiliation{Centre de Recherches sur les tr{\`e}s Basses Temp{\'e}ratures C.N.R.S., BP 166, 38042 Grenoble Cedex 9, France}

\author{N. M. Nemes}
\affiliation{GFMC, Departamento de Física Aplicada III, Campus Moncloa, Universidad Complutense Madrid, ES-28040 Madrid, Spain}

\author{M. Garc{\'i}a-Hern{\'a}ndez}
\affiliation{Instituto de Ciencia de Materiales de Madrid-CSIC, Cantoblanco E-28049 Madrid, Spain}

\author{L. Cario}
\affiliation{Institut des Mat{\'e}riaux Jean Rouxel (IMN), Universit{\'e} de Nantes, CNRS, 2 rue de la Houssini{\`e}re, BP 32229, 44322 Nantes Cedex 03, France}

\author{P. Rodi{\`e}re}
\affiliation{Institut N{\'e}el, CRS/UJF, 25, Av. des Martyrs, BP166, 38042 Grenoble Cedex 9, France}

\author{S. Vieira}
\affiliation{Laboratorio de Bajas Temperaturas, Departamento de F\'isica de la Materia Condensada, Instituto de Ciencia de
Materiales Nicol\'as Cabrera, Facultad de Ciencias \\ Universidad Aut\'onoma de Madrid, E-28049 Madrid, Spain}

\author{H. Suderow}
\affiliation{Laboratorio de Bajas Temperaturas, Departamento de F\'isica de la Materia Condensada, Instituto de Ciencia de
Materiales Nicol\'as Cabrera, Facultad de Ciencias \\ Universidad Aut\'onoma de Madrid, E-28049 Madrid, Spain}

\begin{abstract}
We present measurements of the superconducting critical temperature $T_c$ and upper critical field $H_{c2}$ as a function of pressure in the transition metal dichalcogenide 2H-NbS$_2$ up to $20$ GPa. We observe that $T_c$ increases smoothly from $6$ K at ambient pressure to about $8.9$ K at $20$ GPa. This range of increase is comparable to the one found previously in 2H-NbSe$_2$. The temperature dependence of the upper critical field $H_{c2}(T)$ of 2H-NbS$_2$ varies considerably when increasing the pressure. At low pressures, $H_{c2}(0)$ decreases, and at higher pressures both $T_c$ and $H_{c2}(0)$ increase simultaneously. This points out that there are pressure induced changes of the Fermi surface, which we analyze in terms of a simplified two band approach.
\end{abstract}
 \date{\today} \maketitle

\section{Introduction}

2H-NbSe$_2$ belongs to the family of transition metal dichalcogenide compounds and presents a charge density wave (CDW) below $T_{CDW}=33$ K, which coexists with superconductivity ($T_c=7.2$ K) \cite{Monceau2012,Yokoya2001,Boaknin2003,Rodrigo2004,Kiss2007,Guillamon2008b,Bulaevskii1976,Coleman1988,Sacks1998,Rossnagel2011}. 2H-NbS$_2$ is a related two-band superconductor, with similar $T_c$, and no charge order \cite{Guillamon2008,Kacmarcik2010}. The crystal structure of these materials consists of hexagonal transition metal - chalcogen sandwiches which are coupled through weak van der Waals forces, leading to hexagonal layers with large $c$-axis constant and strongly anisotropic properties. Compressibility is larger along the $c$-axis than in-plane \cite{Kusmartseva2009}. Changes in the electronic properties are produced by altering the lattice constants, using compositional tuning (substitution \cite{Wilson1975}, irradiation \cite{Tsang1975,Mutka1983} or intercalation between layers \cite{Vicent1980,Guillamon2009,Coronado2010}) and applying pressure. Pressure has been shown to lead to an increase of the critical temperature in 2H-NbSe$_2$ with a maximum $T_c$ around $8.5$ K at $10$ GPa. CDW critical temperature decreases to zero at 5 GPa in this material, and shows significant pressure induced modifications in other compounds of the same family \cite{Feng2012,Morosan2006,Sipos2008,Kusmartseva2009,Suderow2004,Suderow2005}. The importance of local strain has been highlighted recently \cite{Soumyanarayanan2012}. In Ref.\cite{Feng2012}, authors apply pressure to 2H-NbSe$_2$ and find that the effective dimensionality of the electronic structure is increased above $4.6$ GPa.

\begin{figure}
	\centering
		\includegraphics[width=0.3\textwidth]{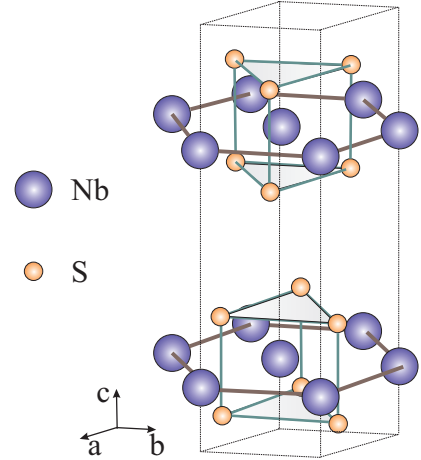}
	\caption{\footnotesize Structure of 2H-NbS$_2$. The niobium atoms are surrounded by sulphur atoms following a trigonal prismatic coordination. The S-Nb-S layers that make up each packet are covalently bound. The coupling between pairs of packets is dominated by van der Waals forces.
	The lattice parameters are $a=3.321{\textrm \AA}$, $b=5.751{\textrm \AA}$ and $c=11.761{\textrm \AA}$ \cite{Fang1995}.}\label{fig:NbS2-lattice}
\end{figure}

To characterize the electronic changes suffered under pressure in bulk materials, the measurement of the upper critical field, $H_{c2}$ is a simple and useful tool. In clean superconductors (with a mean free path, $\ell$, greater than the coherence length, $\xi$), the usual dependence of $H_{c2}$ on temperature is given by Helfand-Werthammer theory, which assumes a simple single band spherical Fermi surface \cite{Helfand1966}. It consists of a linear increase of $H_{c2}$ close to $T_c$, which flattens out at zero temperature. Within this theory, $H_{c2}(0) =\Phi_0\pi \frac{e^{2-\gamma}}{(\hbar^2)}(\frac{T_c}{v_F})^2$ where $\Phi_0$ is the flux quantum, $\gamma \approx 0.577$ Euler's constant and $v_F$ the Fermi velocity. The slope of the linear increase of $H_{c2}$ close to T$_c$, $\frac{dH_{c2}(T)}{dT}(T_c)$ is also proportional to $(\frac{T_c}{v_F})^2$ \cite{Hohenberg1967,Mattheiss1970}. A detailed treatment for complex Fermi surfaces developed in Refs.\cite{Shulga1998,Dahm2003}, shows that the upper critical field can have a strong dependence on temperature, which allows determining the Fermi velocity $v_{F}$ and electron-phonon coupling $\lambda$ parameters on different parts of the Fermi surface.

In the two-gap compounds MgB$_2$, YNi$_2$B$_2$C and 2H-NbSe$_2$, it was found that the ambient pressure $H_{c2}(T)$ has a strong positive curvature close to T$_c$. The associated difference in the Fermi velocities in both bands leads to differing coherence lengths for each band and thus two characteristic features in the upper critical field, giving the positive curvature observed in the experiment \cite{Shulga1998,Suderow2004,Suderow2005,Moshchalkov2009}. In these compounds, the form of such positively curved $H_{c2}(T)$ has a strong dependence as a function of pressure, from which electron phonon coupling and Fermi surface velocities have been obtained. In 2H-NbS$_2$, scanning-tunneling microscopy also revealed the existence of two superconducting gaps \cite{Guillamon2008}. Subsequent heat capacity measurements also show two-gap superconductivity and a small positive curvature of the upper critical field \cite{Kacmarcik2010}. Moreover, the temperature dependence of the superfluid density measured by lower critical field and magnetic penetration depth is very similar to 2H-NbSe$_2$, and is well described by a two gap model\cite{Leroux2012b,Diener2011}. Here, we present new measurements of $T_c$ and $H_{c2}(T)$ as a function of the applied pressure in 2H-NbS$_2$. We observe a smooth increase in $T_c$ as a function of pressure. $H_{c2}(T)$ has a positive curvature at ambient pressure, with an anomalous pressure dependence, which evidences pressure induced changes in the Fermi surface.

\section{Experiment}

We have measured single crystalline samples of 2H-NbS$_2$, with $T_c$ around 6 K and a residual resistivity ratio around $10$. They were grown as described in Ref.\cite{Fisher1980}, and have a hexagonal shape, with lateral dimensions of about $150\;\mu$m and thickness around $30\;\mu$m. We loaded them into a pressure cell made from copper-beryllium alloy, within the sample space delimited by the diamond anvils, which are $0.7$ mm in culet diameter, and a gasket made from NiMo alloy. The diamond anvils were mounted onto a couple of sapphire cylinders inserted into the bores of two Cu-Be pieces. This choice of materials guarantees that the inductive coupling between the coils and the neighboring parts of the cell can be taken as negligible. The gasket had an initial thickness of $300\;\mu$m, which was later reduced to about $60\;\mu$m in its center, after indentation. A small orifice was made by means of arc discharges between the gasket and a molybdenum needle. Its diameter and depth allowed the insertion of the sample as well as
  of some ruby balls that were used to determine the pressure through the ruby fluorescence method \cite{Horn1989}.
Pressure was transmitted by a methanol-ethanol mixture ($4:1$), which has given quasi-hydrostatic conditions up to the pressures of interest in our experiment \cite{Tateiwa2010,Suderow2004,Suderow2005}. The susceptometer is the same already described elsewhere \cite{Suderow2004}. It was designed to obtain the largest signal to-noise ratio, with a pickup coil wounded very close to the sample space. Two primary-secondary coils systems of about $4$ mm in diameter and $2$ mm in height were used. The first one was located surrounding the sample, which was then within the field created by the primary, whereas the secondary acquired the voltage due to any change occurring in sample's properties. The other primary-secondary assembly was glued beside, with no sample inside. The two primaries were connected in series, so the resulting magnetic field was the same in both cases. The secondary coils were connected in series opposition, so the large signals due to the secondary coils themselves could be removed from the beginning. After further compensation by means of an attenuator and a phase shifter the total signal was detected with a lock-in amplifier. For each applied pressure, $T_c$ and $H_{c2}$ were obtained by measuring the magnetic susceptibility as a function of temperature and at different magnetic fields, applied parallel to the $c$-axis. The critical temperature and field were determined by the onset of the superconducting transition curves, defined as the intersection of two tangents, one to the flat portion of the curve above and the second to the steepest variation in the signal below the superconducting transition.

\section{Results}

\begin{figure}
	\centering
		\includegraphics[width=0.5\textwidth]{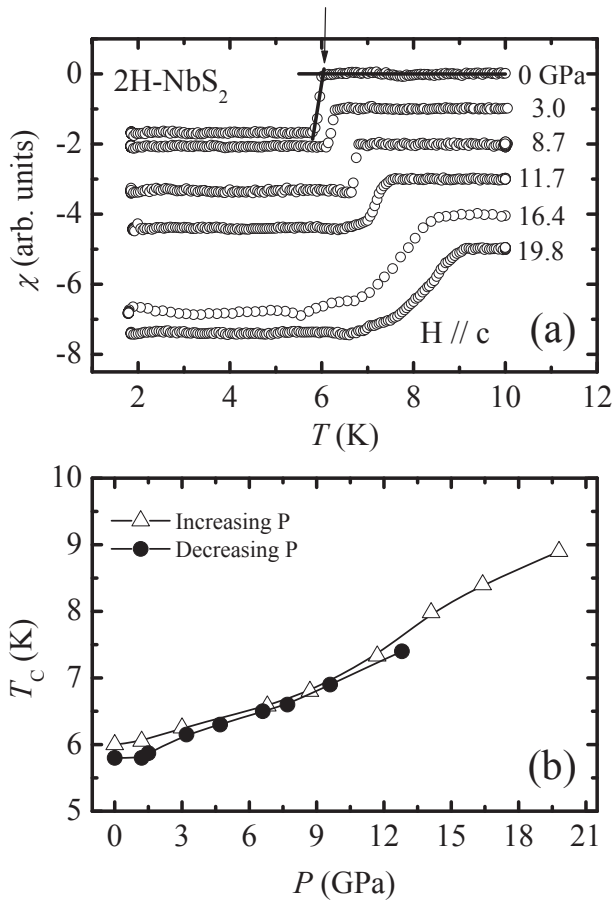}
	\caption{\footnotesize (a) Variation of the susceptibility of a 2H-NbS$_2$ sample as a function of temperature and for several applied pressures. The arrow and the lines show the way we used to extract the corresponding value of $T_c$. (b) Critical temperature as a function of the applied pressure. Triangles and full circles correspond to increasing and decreasing pressure, respectively. Lines are a guide to the eye and simply join data points. Saturation is eventually expected above $20$ GPa.}\label{fig:TcNbS2}
\end{figure}

We find an ambient pressure superconducting critical temperature of $5.7$ K. In previous measurements, no noticeable increase was measured below $1$ GPa \cite{Jones1972,Molinie1974,Barisic2011}. Fig. \ref{fig:TcNbS2}(a) shows the variation with temperature of the susceptibility in 2H-NbS$_2$ sample, for applied pressures ranging between $0$ GPa and $19.8$ GPa. Fig. \ref{fig:TcNbS2}(b) displays the variation of $T_c$ as a function of pressure. Clearly, there is a progressive increment of $T_c$ with pressure. A $T_c$ maximum is likely to exist, but above $20$ GPa. Below $9$ GPa $T_c$ increases with a slope $dT_c/dP= 0.09$ K/GPa, which further grows to $0.22$ K/GPa between $9$ and $14$, and then decreases to $0.16$ K/GPa for higher pressures. 

\begin{figure}
	\centering
		\includegraphics[width=0.5\textwidth]{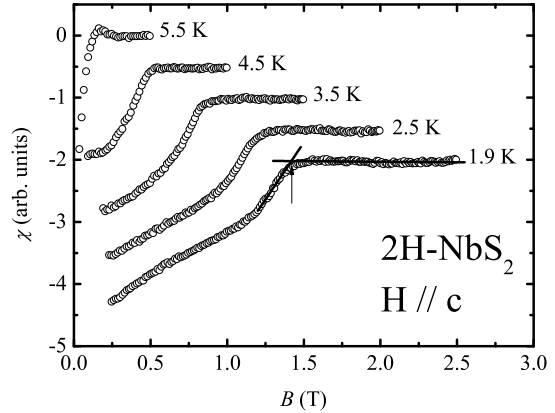}
	\caption{\footnotesize Variation of the susceptibility of a 2H-NbS$_2$ sample as a function of the magnetic field, for a pressure of $3$ GPa and for several temperatures. The arrow and the lines show the way we used to extract the corresponding value of H$_{c2}(T)$. Note that the transition width significantly increases with magnetic field. Other criteria for determining the transition temperature, such as midpoint or lower part of the transition region, lead to similar H$_{c2}(T)$ curves.}\label{fig:Chi}
\end{figure}

The magnetic field dependence of the susceptibility under pressure is shown in Fig.\ref{fig:Chi} for different temperatures and at 3 GPa. There is a smooth evolution of the susceptibility with magnetic field, from which we can easily extract the upper critical field as the onset of the transition. The transition widens significantly at lower temperatures, as expected for type II superconductor. The form of H$_{c2}(T)$, discussed in the following figure, does not depend on chosing H$_{c2}(T)$ from onset, midpoint or lower part of the transition.

\begin{figure}
	\centering
		\includegraphics[width=0.5\textwidth]{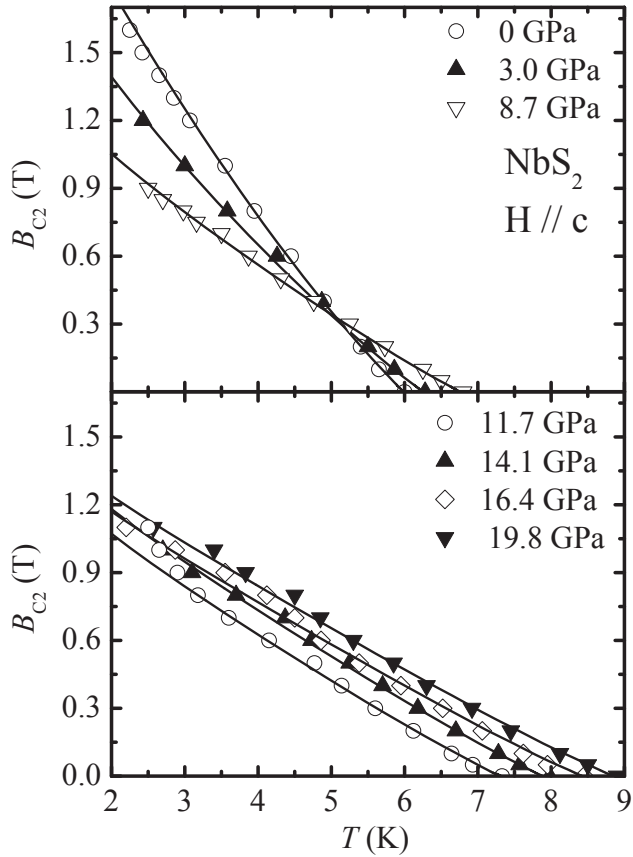}
	\caption{\footnotesize Temperature dependence of the upper critical field H$_{c2}(T)$ as a function of pressure (empty and filled symbols). Fits to the model explained in the text are shown as lines. Note the decrease in the upper critical field until 8.7 GPa (upper panel), which ceases above this pressure (lower panel).}\label{fig:Hc2}
\end{figure}

Previous ambient pressure results on the upper critical field of 2H-NbS$_2$ yielded a positive curvature above $0.5T_c$, and a zero-temperature critical field value of about $2.6-2.7$ T \cite{Kacmarcik2010,Shaw2012}. Our measurements of the ambient pressure temperature dependence of the upper critical field of 2H-NbS$_2$ are shown in the upper panel of Fig.\ref{fig:Hc2}, and essentially confirm previous findings. We observe a slight positive curvature with an extrapolated zero temperature critical field of about $2.6$ T. This positive curvature, although less pronounced, is similar to that found in $H_{c2}(T)$ in 2H-NbSe$_2$. When applying pressure in 2H-NbS$_2$, the critical temperature increases, but the upper critical field at low temperatures decreases strongly up to $8.7$ GPa, above which it increases together with $T_c$. Our results show that $H_{c2}(0)$ drops by a factor $1.5$ between $0$ and $8.7$ GPa and then rises by a factor of 1.5 for highest attained pressures around $20$ GPa.  This is a peculiar behavior. If Fermi surface parameters do not change, theory predicts that $H_{c2}(0)\propto T_c^2$\cite{Helfand1966,Hohenberg1967}.

Thus, the non-monotonous variation of $H_{c2}(T)$ is at odds with most simple single band BCS theory, and we explored how a two band scenario could explain this behaviour. For that purpose, we used the same type of calculations of the upper critical field of strong coupling (multiband) superconductors and its pressure dependence as in \cite{Suderow2004,Suderow2005,Measson2004,Glemot1999}, and numerically linearize $H_{c2}(T)$ equations obtained from Eliashberg theory \cite{Prohammer1987}. The main change of $H_{c2}(T)$ under pressure occurs in the form of the positive curvature and the value of $H_{c2}(0)$, which can be used to determine values of the Fermi surface properties, by parametrizing the Fermi surface in two main different subgroups of electronic excitations. We consider two subgroups of electrons, with bare Fermi velocities $v_{F1}$ and $v_{F2}$, and coupling parameters $\lambda_{ij}$ ($i,j=1,2$, with subindex $1$ for the electronic group with largest pairing strength). There is of course no possibility to deduce a unique set of parameters simply from our $H_{c2}(T,P)$ data:  we rather choose to find the simplest possible scenario which fits the whole pressure dependence of $H_{c2}(T)$.

\begin{figure}
	\centering
		\includegraphics[width=0.5\textwidth]{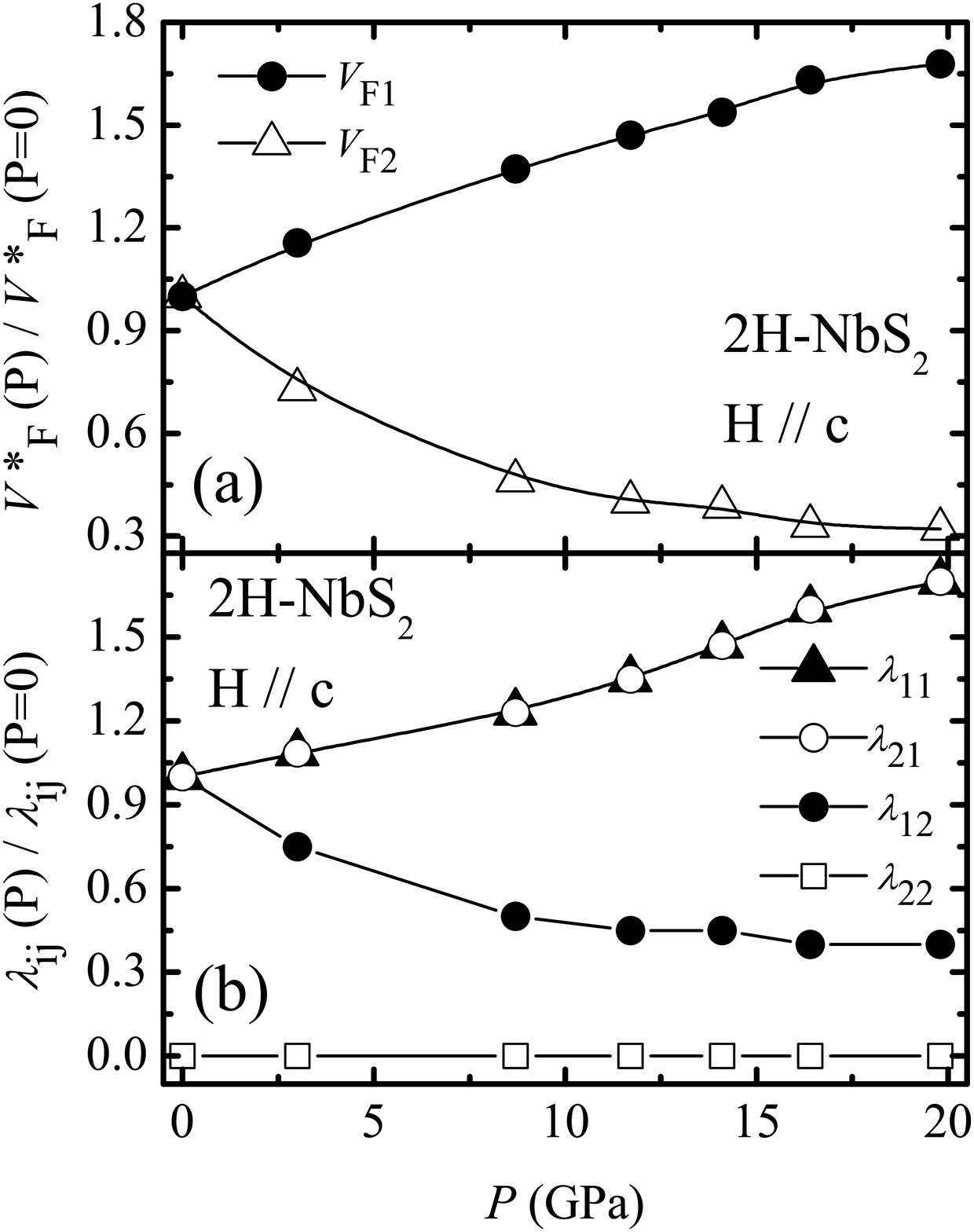}
	\caption{\footnotesize Pressure dependence of the renormalized Fermi surface velocities $v_{Fi}^*(P)/v_{Fi}^*(0)$ (a) and of the electron-phonon coupling parameters $\lambda_{ij}(P)/\lambda_{ij}(0)=\rho_j(P)$ (b) for both subgroups of electrons in 2H-NbS$_2$. Lines simply join points and are a guide to the eyes.}\label{fig:VF-lambda}
\end{figure}

The choice for our scenario was guided by the above mentioned peculiar behavior, namely that $T_c$ increases with pressure, whereas the slope of $H_{c2}(T)$ close to $T_c$ decreases. An increasing $T_c$ suggests an increasing paring strength, but at the same time, the decreasing slope suggests an increased Fermi velocity, and thus, a decreasing effective mass, which is unexpected with increased pairing interactions, or an increased Fermi wave vector. In the latter case, we would also expect an increased bare density of states, for instance through an increased Fermi pocket volume. So we checked if a simple scenario, where the mere increase of the Fermi surface volume of the main electronic group, at the expense of a decreased Fermi surface volume of the second electronic group, could be suitable, with a larger Fermi surface (smaller slope of $H_{c2}$) for the first group to reproduce the change of slope with pressure. To be quantitative, we introduce two (and only two) parameters to describe the pressure evolution of both $T_c$ and $H_{c2}(T)$: these parameters are $\rho_1(P)$ and $\rho_2(P)$, which can be thought as the ratio of the bare density of states of each electronic group under pressure with respect to the density of states at zero pressure, due to the respective Fermi pocket volume change. We can then deduce the pressure evolution of the electron-phonon coupling parameters as: $\lambda_{ij}(P)=\lambda_{ij}(0)\rho_j(P)$, and for the bare Fermi velocities: $v_{Fi}(P)=v_{Fi}(0)\rho_i(P)$. The renormalized Fermi velocities \cite{Prohammer1987} are as usual: $v_{Fi}^*(P)=v_{Fi}(P)\frac{1}{1+\sum_{j}\lambda_{ij}}$. We fixed the values of the mean phonon frequency, $\theta=54.5$ K and Coulomb pseudopotential, $\mu^*=0.1$, and assume that they are pressure independent. We find that indeed, as shown in Fig.\ref{fig:Hc2}, the complete evolution of $H_{c2}(T,P)$ can be fitted when starting from the initial set of values: $\lambda_{11}(0)=1$, $\lambda_{22}(0)=0$, $\lambda_{12}(0)=1.1$, $\lambda_{21}(0)=0.55$, $v_{F1}(0)=3.1$  $10^5 m/s$ ($v_{F1}^*(0)=10^5 m/s$), $v_{F2}(0)=0.155$ $10^5 m/s$ ($v_{F2}^*(0)=10^4 m/s$). With these initial values of the electron phonon coupling constants, pairing is controlled by the first electron group and by the interaction between the first and second electron groups. Fig.\ref{fig:VF-lambda}a shows the pressure evolution of the renormalized Fermi velocities, and Fig.\ref{fig:VF-lambda}b, that of the coupling parameters or equivalently, of $\rho_1$ and $\rho_2$. This demonstrates that $H_{c2}(T,P)$ (including $T_c(P)$) is compatible with a scenario where one of the Fermi pockets expands while the other shrinks, and with a constant pairing potential. Only the bare density of states related to the Fermi pocket volume is changing with pressure.

\section{Discussion and conclusions}

Our results show that the variation of the Fermi surface parameters is monotonic (up to 20GPa), and there are no maxima nor minima as in 2H-NbSe$_2$ \cite{Suderow2005}. In 2H-NbSe$_2$, a dome shape of $T_c(P)$ is found which peaks around $10$ GPa and the CDW disappears at $5$ GPa \cite{Feng2012}, where a kink in $T_c(P)$ is found. The latter is associated to changes in one of the Fermi surfaces, with a possible Lifshitz transition due to one band shifting below the Fermi level. In 2H-NbS$_2$, the strong decrease of the zero temperature upper critical field below $8.7$ GPa, associated with a slight increase in critical temperature, can be explained by a decrease in the renormalized Fermi velocity and an increase in the electron-phonon coupling of one part of the Fermi surface (see Fig. \ref{fig:VF-lambda}), due to Fermi surface modifications. This produces a more pronounced positive curvature in $H_{c2}(T)$, as the differencies in Fermi surface parameters increase. Above about $10$ GPa, variations are smoother and essentially governed by slightly increasing renormalized Fermi velocities of the rest of the Fermi surface.

The Fermi surface of 2H-NbS$_2$ is not known in detail, but has likely the features which are believed to be common to similar transition metal dichalcogenides, namely two pairs of concentric cylindrical sheets derived from Nb 4d electrons \cite{Castro2001,Johannes2006}. Fig. \ref{fig:VF-lambda} shows that pressure induced modifications seem to saturate near $20$ GPa, therefore indicating that $T_c$ cannot be expected to be much higher than $9$ K. It also hints to a shrinking of a part of the Fermi surface, while the other one grows at its expense, as $v_{F1}$ evidences. These features of the superconducting properties cannot be related to a competition with a CDW, as charge order is not present in 2H-NbS$_2$. It has been suggested that the absence of CDW order in this compound is due to anharmonic effects, and that the superconducting properties are essentially determined by the anisotropy and strength of the electron-phonon coupling \cite{Leroux2012}.

2H-NbS$_2$ is at the verge of CDW, which is favored by an increased $a/c$ in other dichalcogenides\cite{Castro2001}. The $a/c$ lattice constant ratio is smaller than in 2H-NbSe$_2$, so that pressure drives farther away from the CDW instability \cite{Castro2001}. This can be fully confirmed, obviously, only by measurements under pressure of any property sensitive to CDW order, such as the resistance. Nevertheless, if CDW re-enters, $T_c(P)$ should show some kink or anomaly, and a decrease instead of the increase we observe here. Thus, our results also show that the re-entrance of CDW is a very unlikely possibility.

It has been postulated that the coexistence of superconductivity with charge density wave is related to the neighborhood to a quantum critical point \cite{Castro2001,Sachdev1999,Coleman2005,Coleman2010}. Quantum critical points appear when a second order phase transition is driven to zero temperature by modifying composition or lattice parameters. Quantum fluctuations diverge at these points and are expected to induce emergent exotic properties \cite{Coleman2010}. In the transition metal dichalcogenides, quantum critical points may appear hidden below the superconducting or charge ordered states \cite{Feng2012}. Pressure in 2H-NbS$_2$  drives the system farther away from such a quantum critical point, which has a marginal effect on the critical temperature, and is not associated to maxima or dome-like shapes of $T_c$. Our data show that maxima in $T_c$ can be reached by going farther away from CDW instability. Therefore, such dome-like shapes in these compounds can be obtained without relation to CDW order.

Finally, let us remark that recent reports discuss synthesis and characterization of single and few layer systems of this and other dichalcogenide compounds. Their superconducting properties, as well as the charge density wave, are expected to differ from the bulk \cite{Frindt1972,Qin1991,Castellanos2010,Augusto2012,Wang2012}. In these systems, the strain induced in the fabrication method is possibly significant and will be of importance in the pairing interaction. The pressure dependence of the bulk properties should be thus useful to predict and understand the modifications found the size reduction down to single or few layers\cite{Calandra2009}.

In summary, we have presented results on the variation of $T_c$ and $H_{c2}$ as a function of pressure in the dichalcogenide 2H-NbS$_2$. Our data show that the critical temperature increases smoothly as pressure is increased. This behavior is compared to the one already found in 2H-NbSe$_2$, where a maximum in T$_c$ is found at 10 GPa. On the other hand, the upper critical field of 2H-NbS$_2$ exhibits an intriguing behavior. There is an initial decrease, contrasting with the increase of $T_c$, but above $8.7$ GPa the upper critical field rises again. We provide a model to explain this behavior in terms of pressure induced changes in the Fermi surface.

The Laboratorio de Bajas Temperaturas is associated to the ICMM of the CSIC. This work was supported by Spanish MINECO (Consolider-Ingenio CSD2007-00010 and CSD2009-00013 programs, SAB2009-0057, FIS2011-23488 and MAT2011-27470-C02-02), by the Comunidad de Madrid through program Nanobiomagnet and by NanoSc-COST program.


\begin{thebibliography}{52}
\expandafter\ifx\csname natexlab\endcsname\relax\def\natexlab#1{#1}\fi
\expandafter\ifx\csname bibnamefont\endcsname\relax
  \def\bibnamefont#1{#1}\fi
\expandafter\ifx\csname bibfnamefont\endcsname\relax
  \def\bibfnamefont#1{#1}\fi
\expandafter\ifx\csname citenamefont\endcsname\relax
  \def\citenamefont#1{#1}\fi
\expandafter\ifx\csname url\endcsname\relax
  \def\url#1{\texttt{#1}}\fi
\expandafter\ifx\csname urlprefix\endcsname\relax\def\urlprefix{URL }\fi
\providecommand{\bibinfo}[2]{#2}
\providecommand{\eprint}[2][]{\url{#2}}

\bibitem[{\citenamefont{Monceau}(2012)}]{Monceau2012}
\bibinfo{author}{\bibfnamefont{P.}~\bibnamefont{Monceau}},
  \bibinfo{journal}{Adv. Phys.} \textbf{\bibinfo{volume}{61}},
  \bibinfo{pages}{325} (\bibinfo{year}{2012}).

\bibitem[{\citenamefont{Yokoya et~al.}(2001)\citenamefont{Yokoya, Kiss,
  Chainani, Shin, Nohara, and Takagi}}]{Yokoya2001}
\bibinfo{author}{\bibfnamefont{T.}~\bibnamefont{Yokoya}},
  \bibinfo{author}{\bibfnamefont{T.}~\bibnamefont{Kiss}},
  \bibinfo{author}{\bibfnamefont{A.}~\bibnamefont{Chainani}},
  \bibinfo{author}{\bibfnamefont{S.}~\bibnamefont{Shin}},
  \bibinfo{author}{\bibfnamefont{M.}~\bibnamefont{Nohara}}, \bibnamefont{and}
  \bibinfo{author}{\bibfnamefont{H.}~\bibnamefont{Takagi}},
  \bibinfo{journal}{Science} \textbf{\bibinfo{volume}{294}},
  \bibinfo{pages}{2518} (\bibinfo{year}{2001}).

\bibitem[{\citenamefont{Boaknin et~al.}(2003)\citenamefont{Boaknin, Tanatar,
  Paglione, Hawthorn, Ronning, Hill, Sutherland, Taillefer, Sonier, Hayden
  et~al.}}]{Boaknin2003}
\bibinfo{author}{\bibfnamefont{E.}~\bibnamefont{Boaknin}},
  \bibinfo{author}{\bibfnamefont{M.~A.} \bibnamefont{Tanatar}},
  \bibinfo{author}{\bibfnamefont{J.}~\bibnamefont{Paglione}},
  \bibinfo{author}{\bibfnamefont{D.}~\bibnamefont{Hawthorn}},
  \bibinfo{author}{\bibfnamefont{F.}~\bibnamefont{Ronning}},
  \bibinfo{author}{\bibfnamefont{R.}~\bibnamefont{Hill}},
  \bibinfo{author}{\bibfnamefont{M.}~\bibnamefont{Sutherland}},
  \bibinfo{author}{\bibfnamefont{L.}~\bibnamefont{Taillefer}},
  \bibinfo{author}{\bibfnamefont{J.}~\bibnamefont{Sonier}},
  \bibinfo{author}{\bibfnamefont{S.}~\bibnamefont{Hayden}},
  \bibnamefont{et~al.}, \bibinfo{journal}{Phys. Rev. Lett.}
  \textbf{\bibinfo{volume}{90}}, \bibinfo{pages}{117003}
  (\bibinfo{year}{2003}).

\bibitem[{\citenamefont{Rodrigo and Vieira}(2004)}]{Rodrigo2004}
\bibinfo{author}{\bibfnamefont{J.~G.} \bibnamefont{Rodrigo}} \bibnamefont{and}
  \bibinfo{author}{\bibfnamefont{S.}~\bibnamefont{Vieira}},
  \bibinfo{journal}{Physica C} \textbf{\bibinfo{volume}{404}},
  \bibinfo{pages}{306} (\bibinfo{year}{2004}).

\bibitem[{\citenamefont{Kiss et~al.}(2007)\citenamefont{Kiss, Yokoya, Chainani,
  Shin, Hanaguri, Nohara, and Takagi}}]{Kiss2007}
\bibinfo{author}{\bibfnamefont{T.}~\bibnamefont{Kiss}},
  \bibinfo{author}{\bibfnamefont{T.}~\bibnamefont{Yokoya}},
  \bibinfo{author}{\bibfnamefont{A.}~\bibnamefont{Chainani}},
  \bibinfo{author}{\bibfnamefont{S.}~\bibnamefont{Shin}},
  \bibinfo{author}{\bibfnamefont{T.}~\bibnamefont{Hanaguri}},
  \bibinfo{author}{\bibfnamefont{M.}~\bibnamefont{Nohara}}, \bibnamefont{and}
  \bibinfo{author}{\bibfnamefont{H.}~\bibnamefont{Takagi}},
  \bibinfo{journal}{Nature Phys.} \textbf{\bibinfo{volume}{3}},
  \bibinfo{pages}{720} (\bibinfo{year}{2007}).

\bibitem[{\citenamefont{Guillamon et~al.}(2008)\citenamefont{Guillamon,
  Suderow, Guinea, and Vieira}}]{Guillamon2008b}
\bibinfo{author}{\bibfnamefont{I.}~\bibnamefont{Guillamon}},
  \bibinfo{author}{\bibfnamefont{H.}~\bibnamefont{Suderow}},
  \bibinfo{author}{\bibfnamefont{F.}~\bibnamefont{Guinea}}, \bibnamefont{and}
  \bibinfo{author}{\bibfnamefont{S.}~\bibnamefont{Vieira}},
  \bibinfo{journal}{Phys. Rev. B} \textbf{\bibinfo{volume}{77}},
  \bibinfo{pages}{134505} (\bibinfo{year}{2008}).

\bibitem[{\citenamefont{Bulaevski{\u i}}(1976)}]{Bulaevskii1976}
\bibinfo{author}{\bibfnamefont{L.~N.} \bibnamefont{Bulaevski{\u i}}},
  \bibinfo{journal}{Sov. Phys. Usp.} \textbf{\bibinfo{volume}{19}},
  \bibinfo{pages}{836} (\bibinfo{year}{1976}).

\bibitem[{\citenamefont{Coleman et~al.}(1988)\citenamefont{Coleman,
  Giambattista, Hansma, Johnson, McNairy, and Slough}}]{Coleman1988}
\bibinfo{author}{\bibfnamefont{R.~V.} \bibnamefont{Coleman}},
  \bibinfo{author}{\bibfnamefont{B.}~\bibnamefont{Giambattista}},
  \bibinfo{author}{\bibfnamefont{P.~K.} \bibnamefont{Hansma}},
  \bibinfo{author}{\bibfnamefont{A.}~\bibnamefont{Johnson}},
  \bibinfo{author}{\bibfnamefont{W.~W.} \bibnamefont{McNairy}},
  \bibnamefont{and} \bibinfo{author}{\bibfnamefont{C.~G.}
  \bibnamefont{Slough}}, \bibinfo{journal}{Adv. Phys.}
  \textbf{\bibinfo{volume}{37}}, \bibinfo{pages}{559} (\bibinfo{year}{1988}).

\bibitem[{\citenamefont{Sacks et~al.}(1998)\citenamefont{Sacks, Roditchev, and
  Klein}}]{Sacks1998}
\bibinfo{author}{\bibfnamefont{W.}~\bibnamefont{Sacks}},
  \bibinfo{author}{\bibfnamefont{D.}~\bibnamefont{Roditchev}},
  \bibnamefont{and} \bibinfo{author}{\bibfnamefont{J.}~\bibnamefont{Klein}},
  \bibinfo{journal}{Phys. Rev. B} \textbf{\bibinfo{volume}{57}},
  \bibinfo{pages}{13118} (\bibinfo{year}{1998}).

\bibitem[{\citenamefont{Rossnagel}(2011)}]{Rossnagel2011}
\bibinfo{author}{\bibfnamefont{K.}~\bibnamefont{Rossnagel}},
  \bibinfo{journal}{J. Phys.: Condens. Matter} \textbf{\bibinfo{volume}{23}},
  \bibinfo{pages}{213001} (\bibinfo{year}{2011}).

\bibitem[{\citenamefont{Guillam{\'o}n et~al.}(2008)\citenamefont{Guillam{\'o}n,
  Suderow, Vieira, Cario, Diener, and Rodi{\`e}re}}]{Guillamon2008}
\bibinfo{author}{\bibfnamefont{I.}~\bibnamefont{Guillam{\'o}n}},
  \bibinfo{author}{\bibfnamefont{H.}~\bibnamefont{Suderow}},
  \bibinfo{author}{\bibfnamefont{S.}~\bibnamefont{Vieira}},
  \bibinfo{author}{\bibfnamefont{L.}~\bibnamefont{Cario}},
  \bibinfo{author}{\bibfnamefont{P.}~\bibnamefont{Diener}}, \bibnamefont{and}
  \bibinfo{author}{\bibfnamefont{P.}~\bibnamefont{Rodi{\`e}re}},
  \bibinfo{journal}{Phys. Rev. Lett.} \textbf{\bibinfo{volume}{101}},
  \bibinfo{pages}{166407} (\bibinfo{year}{2008}).

\bibitem[{\citenamefont{Ka{\v c}mar{\v c}ik et~al.}(2010)\citenamefont{Ka{\v
  c}mar{\v c}ik, Pribulov{\'a}, Marcenat, Klein, Rodi{\`e}re, and
  Samuely}}]{Kacmarcik2010}
\bibinfo{author}{\bibfnamefont{J.}~\bibnamefont{Ka{\v c}mar{\v c}ik}},
  \bibinfo{author}{\bibfnamefont{Z.}~\bibnamefont{Pribulov{\'a}}},
  \bibinfo{author}{\bibfnamefont{C.}~\bibnamefont{Marcenat}},
  \bibinfo{author}{\bibfnamefont{T.}~\bibnamefont{Klein}},
  \bibinfo{author}{\bibfnamefont{P.}~\bibnamefont{Rodi{\`e}re}},
  \bibnamefont{and} \bibinfo{author}{\bibfnamefont{L.~C.~P.}
  \bibnamefont{Samuely}}, \bibinfo{journal}{Phys. Rev. B}
  \textbf{\bibinfo{volume}{82}}, \bibinfo{pages}{014518}
  (\bibinfo{year}{2010}).

\bibitem[{\citenamefont{Kusmartseva et~al.}(2009)\citenamefont{Kusmartseva,
  Sipos, Berger, Forr{\'o}, and Tuti{\v{s}}}}]{Kusmartseva2009}
\bibinfo{author}{\bibfnamefont{A.~F.} \bibnamefont{Kusmartseva}},
  \bibinfo{author}{\bibfnamefont{B.}~\bibnamefont{Sipos}},
  \bibinfo{author}{\bibfnamefont{H.}~\bibnamefont{Berger}},
  \bibinfo{author}{\bibfnamefont{L.}~\bibnamefont{Forr{\'o}}},
  \bibnamefont{and}
  \bibinfo{author}{\bibfnamefont{E.}~\bibnamefont{Tuti{\v{s}}}},
  \bibinfo{journal}{Phys. Rev. Lett.} \textbf{\bibinfo{volume}{103}},
  \bibinfo{pages}{236401} (\bibinfo{year}{2009}).

\bibitem[{\citenamefont{Wilson et~al.}(1975)\citenamefont{Wilson, DiSalvo, and
  Mahajan}}]{Wilson1975}
\bibinfo{author}{\bibfnamefont{J.~A.} \bibnamefont{Wilson}},
  \bibinfo{author}{\bibfnamefont{F.~J.} \bibnamefont{DiSalvo}},
  \bibnamefont{and} \bibinfo{author}{\bibfnamefont{S.}~\bibnamefont{Mahajan}},
  \bibinfo{journal}{Adv. Phys.} \textbf{\bibinfo{volume}{24}},
  \bibinfo{pages}{117} (\bibinfo{year}{1975}).

\bibitem[{\citenamefont{Tsang et~al.}(1975)\citenamefont{Tsang, Shafer, and
  Crowder}}]{Tsang1975}
\bibinfo{author}{\bibfnamefont{J.~C.} \bibnamefont{Tsang}},
  \bibinfo{author}{\bibfnamefont{M.~W.} \bibnamefont{Shafer}},
  \bibnamefont{and} \bibinfo{author}{\bibfnamefont{B.~L.}
  \bibnamefont{Crowder}}, \bibinfo{journal}{Phys. Rev. B}
  \textbf{\bibinfo{volume}{11}}, \bibinfo{pages}{155} (\bibinfo{year}{1975}).

\bibitem[{\citenamefont{Mutka}(1983)}]{Mutka1983}
\bibinfo{author}{\bibfnamefont{H.}~\bibnamefont{Mutka}},
  \bibinfo{journal}{Phys. Rev. B} \textbf{\bibinfo{volume}{28}},
  \bibinfo{pages}{2855} (\bibinfo{year}{1983}).

\bibitem[{\citenamefont{Vicent et~al.}(1980)\citenamefont{Vicent, Hillenius,
  and Coleman}}]{Vicent1980}
\bibinfo{author}{\bibfnamefont{J.~L.} \bibnamefont{Vicent}},
  \bibinfo{author}{\bibfnamefont{S.~J.} \bibnamefont{Hillenius}},
  \bibnamefont{and} \bibinfo{author}{\bibfnamefont{H.~V.}
  \bibnamefont{Coleman}}, \bibinfo{journal}{Phys. Rev. Lett.}
  \textbf{\bibinfo{volume}{44}}, \bibinfo{pages}{892} (\bibinfo{year}{1980}).

\bibitem[{\citenamefont{Guillam{\'o}n et~al.}(2009)\citenamefont{Guillam{\'o}n,
  Suderow, Fern{\'a}ndez-Pacheco, Ses{\'e}, C{\'o}rdoba, Teresa, Ibarra, and
  Vieira}}]{Guillamon2009}
\bibinfo{author}{\bibfnamefont{I.}~\bibnamefont{Guillam{\'o}n}},
  \bibinfo{author}{\bibfnamefont{H.}~\bibnamefont{Suderow}},
  \bibinfo{author}{\bibfnamefont{A.}~\bibnamefont{Fern{\'a}ndez-Pacheco}},
  \bibinfo{author}{\bibfnamefont{J.}~\bibnamefont{Ses{\'e}}},
  \bibinfo{author}{\bibfnamefont{R.}~\bibnamefont{C{\'o}rdoba}},
  \bibinfo{author}{\bibfnamefont{J.~M.~D.} \bibnamefont{Teresa}},
  \bibinfo{author}{\bibfnamefont{M.~R.} \bibnamefont{Ibarra}},
  \bibnamefont{and} \bibinfo{author}{\bibfnamefont{S.}~\bibnamefont{Vieira}},
  \bibinfo{journal}{Nat. Phys.} \textbf{\bibinfo{volume}{5}},
  \bibinfo{pages}{651} (\bibinfo{year}{2009}).

\bibitem[{\citenamefont{Coronado et~al.}(2010)\citenamefont{Coronado,
  Mart{\'i}-Gastaldo, Navarro-Moratalla, Ribera, Blundell, and
  Baker}}]{Coronado2010}
\bibinfo{author}{\bibfnamefont{E.}~\bibnamefont{Coronado}},
  \bibinfo{author}{\bibfnamefont{C.}~\bibnamefont{Mart{\'i}-Gastaldo}},
  \bibinfo{author}{\bibfnamefont{E.}~\bibnamefont{Navarro-Moratalla}},
  \bibinfo{author}{\bibfnamefont{A.}~\bibnamefont{Ribera}},
  \bibinfo{author}{\bibfnamefont{S.~J.} \bibnamefont{Blundell}},
  \bibnamefont{and} \bibinfo{author}{\bibfnamefont{P.~J.} \bibnamefont{Baker}},
  \bibinfo{journal}{Nat. Chem.} \textbf{\bibinfo{volume}{2}},
  \bibinfo{pages}{1031} (\bibinfo{year}{2010}).

\bibitem[{\citenamefont{Feng et~al.}(2012)\citenamefont{Feng, Wang, Jaramillo,
  van Wezel, Haravifard, Srajer, Liu, Xu, Littlewood, and
  Rosenbaum}}]{Feng2012}
\bibinfo{author}{\bibfnamefont{Y.}~\bibnamefont{Feng}},
  \bibinfo{author}{\bibfnamefont{J.}~\bibnamefont{Wang}},
  \bibinfo{author}{\bibfnamefont{R.}~\bibnamefont{Jaramillo}},
  \bibinfo{author}{\bibfnamefont{J.}~\bibnamefont{van Wezel}},
  \bibinfo{author}{\bibfnamefont{S.}~\bibnamefont{Haravifard}},
  \bibinfo{author}{\bibfnamefont{G.}~\bibnamefont{Srajer}},
  \bibinfo{author}{\bibfnamefont{Y.}~\bibnamefont{Liu}},
  \bibinfo{author}{\bibfnamefont{Z.-A.} \bibnamefont{Xu}},
  \bibinfo{author}{\bibfnamefont{P.~B.} \bibnamefont{Littlewood}},
  \bibnamefont{and} \bibinfo{author}{\bibfnamefont{T.~F.}
  \bibnamefont{Rosenbaum}}, \bibinfo{journal}{PNAS}
  \textbf{\bibinfo{volume}{109}}, \bibinfo{pages}{7224} (\bibinfo{year}{2012}).

\bibitem[{\citenamefont{Morosan et~al.}(2006)\citenamefont{Morosan, Zandbergen,
  Dennis, Bos, Onose, Klimczuk, Ramirez, Ong, and Cava}}]{Morosan2006}
\bibinfo{author}{\bibfnamefont{E.}~\bibnamefont{Morosan}},
  \bibinfo{author}{\bibfnamefont{H.~W.} \bibnamefont{Zandbergen}},
  \bibinfo{author}{\bibfnamefont{B.~S.} \bibnamefont{Dennis}},
  \bibinfo{author}{\bibfnamefont{J.~W.~G.} \bibnamefont{Bos}},
  \bibinfo{author}{\bibfnamefont{Y.}~\bibnamefont{Onose}},
  \bibinfo{author}{\bibfnamefont{T.}~\bibnamefont{Klimczuk}},
  \bibinfo{author}{\bibfnamefont{A.~P.} \bibnamefont{Ramirez}},
  \bibinfo{author}{\bibfnamefont{N.~P.} \bibnamefont{Ong}}, \bibnamefont{and}
  \bibinfo{author}{\bibfnamefont{R.~J.} \bibnamefont{Cava}},
  \bibinfo{journal}{Nat. Phys.} \textbf{\bibinfo{volume}{2}},
  \bibinfo{pages}{544} (\bibinfo{year}{2006}).

\bibitem[{\citenamefont{Sipos et~al.}(2008)\citenamefont{Sipos, Kusmartseva,
  Akrap, Berger, Forro, and Tutis}}]{Sipos2008}
\bibinfo{author}{\bibfnamefont{B.}~\bibnamefont{Sipos}},
  \bibinfo{author}{\bibfnamefont{A.~F.} \bibnamefont{Kusmartseva}},
  \bibinfo{author}{\bibfnamefont{A.}~\bibnamefont{Akrap}},
  \bibinfo{author}{\bibfnamefont{H.}~\bibnamefont{Berger}},
  \bibinfo{author}{\bibfnamefont{L.}~\bibnamefont{Forro}}, \bibnamefont{and}
  \bibinfo{author}{\bibfnamefont{E.}~\bibnamefont{Tutis}},
  \bibinfo{journal}{Nat. Mater.} \textbf{\bibinfo{volume}{7}},
  \bibinfo{pages}{960} (\bibinfo{year}{2008}).

\bibitem[{\citenamefont{Suderow et~al.}(2004)\citenamefont{Suderow, Tissen,
  Brison, Mart{\'i}nez, Vieira, Lejay, Lee, and Tajima}}]{Suderow2004}
\bibinfo{author}{\bibfnamefont{H.}~\bibnamefont{Suderow}},
  \bibinfo{author}{\bibfnamefont{V.~G.} \bibnamefont{Tissen}},
  \bibinfo{author}{\bibfnamefont{J.~P.} \bibnamefont{Brison}},
  \bibinfo{author}{\bibfnamefont{J.~L.} \bibnamefont{Mart{\'i}nez}},
  \bibinfo{author}{\bibfnamefont{S.}~\bibnamefont{Vieira}},
  \bibinfo{author}{\bibfnamefont{P.}~\bibnamefont{Lejay}},
  \bibinfo{author}{\bibfnamefont{S.}~\bibnamefont{Lee}}, \bibnamefont{and}
  \bibinfo{author}{\bibfnamefont{S.}~\bibnamefont{Tajima}},
  \bibinfo{journal}{Phys. Rev. B} \textbf{\bibinfo{volume}{70}},
  \bibinfo{pages}{134518} (\bibinfo{year}{2004}).

\bibitem[{\citenamefont{Suderow et~al.}(2005)\citenamefont{Suderow, Tissen,
  Brison, Mart{\'i}nez, and Vieira}}]{Suderow2005}
\bibinfo{author}{\bibfnamefont{H.}~\bibnamefont{Suderow}},
  \bibinfo{author}{\bibfnamefont{V.~G.} \bibnamefont{Tissen}},
  \bibinfo{author}{\bibfnamefont{J.~P.} \bibnamefont{Brison}},
  \bibinfo{author}{\bibfnamefont{J.~L.} \bibnamefont{Mart{\'i}nez}},
  \bibnamefont{and} \bibinfo{author}{\bibfnamefont{S.}~\bibnamefont{Vieira}},
  \bibinfo{journal}{Phys. Rev. Lett.} \textbf{\bibinfo{volume}{95}},
  \bibinfo{pages}{117006} (\bibinfo{year}{2005}).

\bibitem[{\citenamefont{Soumyanarayanan
  et~al.}(2012)\citenamefont{Soumyanarayanan, Yee, He, van Wezel, Rahn,
  Rossnagel, Hudson, Norman, and Hoffman}}]{Soumyanarayanan2012}
\bibinfo{author}{\bibfnamefont{A.}~\bibnamefont{Soumyanarayanan}},
  \bibinfo{author}{\bibfnamefont{M.~M.} \bibnamefont{Yee}},
  \bibinfo{author}{\bibfnamefont{Y.}~\bibnamefont{He}},
  \bibinfo{author}{\bibfnamefont{J.}~\bibnamefont{van Wezel}},
  \bibinfo{author}{\bibfnamefont{D.~J.} \bibnamefont{Rahn}},
  \bibinfo{author}{\bibfnamefont{K.}~\bibnamefont{Rossnagel}},
  \bibinfo{author}{\bibfnamefont{E.~W.} \bibnamefont{Hudson}},
  \bibinfo{author}{\bibfnamefont{M.~R.} \bibnamefont{Norman}},
  \bibnamefont{and} \bibinfo{author}{\bibfnamefont{J.~E.}
  \bibnamefont{Hoffman}}, \bibinfo{journal}{arXiv:1212.4087 [cond-mat.str-el]}
  (\bibinfo{year}{2012}).

\bibitem[{\citenamefont{Fang et~al.}(1995)\citenamefont{Fang, Ettema, Haas,
  Wiegers, van Leuken, and de~Groot}}]{Fang1995}
\bibinfo{author}{\bibfnamefont{C.~M.} \bibnamefont{Fang}},
  \bibinfo{author}{\bibfnamefont{A.~R. H.~F.} \bibnamefont{Ettema}},
  \bibinfo{author}{\bibfnamefont{C.}~\bibnamefont{Haas}},
  \bibinfo{author}{\bibfnamefont{G.~A.} \bibnamefont{Wiegers}},
  \bibinfo{author}{\bibfnamefont{H.}~\bibnamefont{van Leuken}},
  \bibnamefont{and} \bibinfo{author}{\bibfnamefont{R.~A.}
  \bibnamefont{de~Groot}}, \bibinfo{journal}{Phys. Rev. B}
  \textbf{\bibinfo{volume}{52}}, \bibinfo{pages}{2336} (\bibinfo{year}{1995}).

\bibitem[{\citenamefont{Helfand and Werthamer}(1966)}]{Helfand1966}
\bibinfo{author}{\bibfnamefont{E.}~\bibnamefont{Helfand}} \bibnamefont{and}
  \bibinfo{author}{\bibfnamefont{N.}~\bibnamefont{Werthamer}},
  \bibinfo{journal}{Phys. Rev.} \textbf{\bibinfo{volume}{147}},
  \bibinfo{pages}{288} (\bibinfo{year}{1966}).

\bibitem[{\citenamefont{Hohenberg et~al.}(1967)\citenamefont{Hohenberg and Werthamer}}]{Hohenberg1967}
\bibinfo{author}{\bibfnamefont{P.~C.} \bibnamefont{Hohenberg}},
  \bibinfo{author}{\bibfnamefont{N.~R.}~\bibnamefont{Werthamer}},
\bibinfo{journal}{Physical Review} \textbf{\bibinfo{volume}{153}}, \bibinfo{pages}{493} (\bibinfo{year}{1967}).

\bibitem[{\citenamefont{Mattheiss}(1970)\citenamefont{Mattheiss}}]{Mattheiss1970}
\bibinfo{author}{\bibfnamefont{L.~F.} \bibnamefont{Mattheiss}},
\bibinfo{journal}{Physical Review B} \textbf{\bibinfo{volume}{1}}, \bibinfo{pages}{373} (\bibinfo{year}{1970}).

\bibitem[{\citenamefont{Shulga et~al.}(1998)\citenamefont{Shulga, Drechsler,
  Fuchs, M{\"u}ller, Winzer, Heinecke, and Krug}}]{Shulga1998}
\bibinfo{author}{\bibfnamefont{S.~V.} \bibnamefont{Shulga}},
  \bibinfo{author}{\bibfnamefont{S.~L.} \bibnamefont{Drechsler}},
  \bibinfo{author}{\bibfnamefont{G.}~\bibnamefont{Fuchs}},
  \bibinfo{author}{\bibfnamefont{K.~H.} \bibnamefont{M{\"u}ller}},
  \bibinfo{author}{\bibfnamefont{K.}~\bibnamefont{Winzer}},
  \bibinfo{author}{\bibfnamefont{M.}~\bibnamefont{Heinecke}}, \bibnamefont{and}
  \bibinfo{author}{\bibfnamefont{K.}~\bibnamefont{Krug}},
  \bibinfo{journal}{Phys. Rev. Lett.} \textbf{\bibinfo{volume}{80}},
  \bibinfo{pages}{1730} (\bibinfo{year}{1998}).

\bibitem[{\citenamefont{Dahm and Schopohl}(2003)}]{Dahm2003}
\bibinfo{author}{\bibfnamefont{T.}~\bibnamefont{Dahm}} \bibnamefont{and}
  \bibinfo{author}{\bibfnamefont{N.}~\bibnamefont{Schopohl}},
  \bibinfo{journal}{Phys. Rev. Lett.} \textbf{\bibinfo{volume}{91}},
  \bibinfo{pages}{017001} (\bibinfo{year}{2003}).

\bibitem[{\citenamefont{Moshchalkov et~al.}(2009)\citenamefont{Moshchalkov,
  Menghini, Nishio, Chen, Silhanek, Dao, Chibotaru, Zhigadlo, and
  Karpinski}}]{Moshchalkov2009}
\bibinfo{author}{\bibfnamefont{V.}~\bibnamefont{Moshchalkov}},
  \bibinfo{author}{\bibfnamefont{M.}~\bibnamefont{Menghini}},
  \bibinfo{author}{\bibfnamefont{T.}~\bibnamefont{Nishio}},
  \bibinfo{author}{\bibfnamefont{Q.}~\bibnamefont{Chen}},
  \bibinfo{author}{\bibfnamefont{A.}~\bibnamefont{Silhanek}},
  \bibinfo{author}{\bibfnamefont{V.}~\bibnamefont{Dao}},
  \bibinfo{author}{\bibfnamefont{L.}~\bibnamefont{Chibotaru}},
  \bibinfo{author}{\bibfnamefont{N.}~\bibnamefont{Zhigadlo}}, \bibnamefont{and}
  \bibinfo{author}{\bibfnamefont{J.}~\bibnamefont{Karpinski}},
  \bibinfo{journal}{Phys. Rev. Lett.} \textbf{\bibinfo{volume}{102}},
  \bibinfo{pages}{117001} (\bibinfo{year}{2009}).

\bibitem[{\citenamefont{Leroux et~al.}(2012{\natexlab{a}})\citenamefont{Leroux,
  Rodiere, Cario, and Klein}}]{Leroux2012b}
\bibinfo{author}{\bibfnamefont{M.}~\bibnamefont{Leroux}},
  \bibinfo{author}{\bibfnamefont{P.}~\bibnamefont{Rodiere}},
  \bibinfo{author}{\bibfnamefont{L.}~\bibnamefont{Cario}}, \bibnamefont{and}
  \bibinfo{author}{\bibfnamefont{T.}~\bibnamefont{Klein}},
  \bibinfo{journal}{Physica B Cond Matt} \textbf{\bibinfo{volume}{407}},
  \bibinfo{pages}{1813} (\bibinfo{year}{2012}{\natexlab{a}}).

\bibitem[{\citenamefont{Diener et~al.}(2011)\citenamefont{Diener, Leroux,
  Cario, Klein, and Rodi{\`e}re}}]{Diener2011}
\bibinfo{author}{\bibfnamefont{P.}~\bibnamefont{Diener}},
  \bibinfo{author}{\bibfnamefont{M.}~\bibnamefont{Leroux}},
  \bibinfo{author}{\bibfnamefont{L.}~\bibnamefont{Cario}},
  \bibinfo{author}{\bibfnamefont{T.}~\bibnamefont{Klein}}, \bibnamefont{and}
  \bibinfo{author}{\bibfnamefont{P.}~\bibnamefont{Rodi{\`e}re}},
  \bibinfo{journal}{Phys. Rev. B} \textbf{\bibinfo{volume}{84}},
  \bibinfo{pages}{054531} (\bibinfo{year}{2011}).

\bibitem[{\citenamefont{Fisher and Sienko}(2009)\citenamefont{Fisher, Sienko}}]{Fisher1980}
\bibinfo{author}{\bibfnamefont{W.} \bibnamefont{Fisher}},
  \bibinfo{author}{\bibfnamefont{M.}~\bibnamefont{Sienko}}, \bibinfo{journal}{Inorg. Chem.} \textbf{\bibinfo{volume}{19}}, \bibinfo{pages}{39} (\bibinfo{year}{1980}).

\bibitem[{\citenamefont{Horn and Gupta}(1989)}]{Horn1989}
\bibinfo{author}{\bibfnamefont{P.~D.} \bibnamefont{Horn}} \bibnamefont{and}
  \bibinfo{author}{\bibfnamefont{Y.~M.} \bibnamefont{Gupta}},
  \bibinfo{journal}{Phys. Rev. B} \textbf{\bibinfo{volume}{39}},
  \bibinfo{pages}{973} (\bibinfo{year}{1989}).

\bibitem[{\citenamefont{Tateiwa and Haga}(2010)}]{Tateiwa2010}
\bibinfo{author}{\bibfnamefont{N.}~\bibnamefont{Tateiwa}} \bibnamefont{and}
  \bibinfo{author}{\bibfnamefont{Y.}~\bibnamefont{Haga}}, \bibinfo{journal}{J.
  Phys.: Conf. Series} \textbf{\bibinfo{volume}{215}}, \bibinfo{pages}{012178}
  (\bibinfo{year}{2010}).

\bibitem[{\citenamefont{Jr. et~al.}(1972)\citenamefont{Jr., Shanks, Finnemore,
  and Morosin}}]{Jones1972}
\bibinfo{author}{\bibfnamefont{B.~E.~J.} \bibnamefont{Jr.}},
  \bibinfo{author}{\bibfnamefont{H.~H.} \bibnamefont{Shanks}},
  \bibinfo{author}{\bibfnamefont{D.~K.} \bibnamefont{Finnemore}},
  \bibnamefont{and} \bibinfo{author}{\bibfnamefont{B.}~\bibnamefont{Morosin}},
  \bibinfo{journal}{Phys. Rev. B} \textbf{\bibinfo{volume}{6}},
  \bibinfo{pages}{835} (\bibinfo{year}{1972}).

\bibitem[{\citenamefont{Molini{\'e} et~al.}(1974)\citenamefont{Molini{\'e},
  Jerome, and Grant}}]{Molinie1974}
\bibinfo{author}{\bibfnamefont{P.}~\bibnamefont{Molini{\'e}}},
  \bibinfo{author}{\bibfnamefont{D.}~\bibnamefont{Jerome}}, \bibnamefont{and}
  \bibinfo{author}{\bibfnamefont{A.~J.} \bibnamefont{Grant}},
  \bibinfo{journal}{Phil. Mag.} p. \bibinfo{pages}{1091}
  (\bibinfo{year}{1974}).

\bibitem[{\citenamefont{Bari{\v s}i{\'c} et~al.}(2011)\citenamefont{Bari{\v
  s}i{\'c}, Smiljani{\'c}, Popcevi{\'c}, Bilu{\v s}i{\'c}, Tuti{\v s},
  Smontara, Berger, Ja{\'c}imovi{\'c}, Yuli, and Forr{\'o}}}]{Barisic2011}
\bibinfo{author}{\bibfnamefont{N.}~\bibnamefont{Bari{\v s}i{\'c}}},
  \bibinfo{author}{\bibfnamefont{I.}~\bibnamefont{Smiljani{\'c}}},
  \bibinfo{author}{\bibfnamefont{P.}~\bibnamefont{Popcevi{\'c}}},
  \bibinfo{author}{\bibfnamefont{A.}~\bibnamefont{Bilu{\v s}i{\'c}}},
  \bibinfo{author}{\bibfnamefont{E.}~\bibnamefont{Tuti{\v s}}},
  \bibinfo{author}{\bibfnamefont{A.}~\bibnamefont{Smontara}},
  \bibinfo{author}{\bibfnamefont{H.}~\bibnamefont{Berger}},
  \bibinfo{author}{\bibfnamefont{J.}~\bibnamefont{Ja{\'c}imovi{\'c}}},
  \bibinfo{author}{\bibfnamefont{O.}~\bibnamefont{Yuli}}, \bibnamefont{and}
  \bibinfo{author}{\bibfnamefont{L.}~\bibnamefont{Forr{\'o}}},
  \bibinfo{journal}{Phis. Rev. B} \textbf{\bibinfo{volume}{84}},
  \bibinfo{pages}{075157} (\bibinfo{year}{2011}).

\bibitem[{\citenamefont{Shaw et~al.}(2012)\citenamefont{Shaw, Mandal, Banerjee,
  Niazi, Rastogi, Sood, Ramakrishnan, and Grover}}]{Shaw2012}
\bibinfo{author}{\bibfnamefont{G.}~\bibnamefont{Shaw}},
  \bibinfo{author}{\bibfnamefont{P.}~\bibnamefont{Mandal}},
  \bibinfo{author}{\bibfnamefont{S.~S.} \bibnamefont{Banerjee}},
  \bibinfo{author}{\bibfnamefont{A.}~\bibnamefont{Niazi}},
  \bibinfo{author}{\bibfnamefont{A.~K.} \bibnamefont{Rastogi}},
  \bibinfo{author}{\bibfnamefont{A.~K.} \bibnamefont{Sood}},
  \bibinfo{author}{\bibfnamefont{S.}~\bibnamefont{Ramakrishnan}},
  \bibnamefont{and} \bibinfo{author}{\bibfnamefont{A.~K.}
  \bibnamefont{Grover}}, \bibinfo{journal}{Phys. Rev. B}
  \textbf{\bibinfo{volume}{85}}, \bibinfo{pages}{174517}
  (\bibinfo{year}{2012}).

\bibitem[{\citenamefont{M{\'e}asson et~al.}(2004)\citenamefont{M{\'e}asson,
  Braithwaite, Flouquet, Seyfarth, Brison, Lhotel, Paulsen, Sugawara, and
  Sato}}]{Measson2004}
\bibinfo{author}{\bibfnamefont{M.~A.} \bibnamefont{M{\'e}asson}},
  \bibinfo{author}{\bibfnamefont{D.}~\bibnamefont{Braithwaite}},
  \bibinfo{author}{\bibfnamefont{J.}~\bibnamefont{Flouquet}},
  \bibinfo{author}{\bibfnamefont{G.}~\bibnamefont{Seyfarth}},
  \bibinfo{author}{\bibfnamefont{J.~P.} \bibnamefont{Brison}},
  \bibinfo{author}{\bibfnamefont{E.}~\bibnamefont{Lhotel}},
  \bibinfo{author}{\bibfnamefont{C.}~\bibnamefont{Paulsen}},
  \bibinfo{author}{\bibfnamefont{H.}~\bibnamefont{Sugawara}}, \bibnamefont{and}
  \bibinfo{author}{\bibfnamefont{H.}~\bibnamefont{Sato}},
  \bibinfo{journal}{Phys. Rev. B} \textbf{\bibinfo{volume}{70}},
  \bibinfo{pages}{064516} (\bibinfo{year}{2004}).

\bibitem[{\citenamefont{Glemot et~al.}(1999)\citenamefont{Glemot, Brison,
  Flouquet, Buzdin, Sheikin, Jaccard, Thessieu, and Thomas}}]{Glemot1999}
\bibinfo{author}{\bibfnamefont{L.}~\bibnamefont{Glemot}},
  \bibinfo{author}{\bibfnamefont{J.}~\bibnamefont{Brison}},
  \bibinfo{author}{\bibfnamefont{J.}~\bibnamefont{Flouquet}},
  \bibinfo{author}{\bibfnamefont{A.}~\bibnamefont{Buzdin}},
  \bibinfo{author}{\bibfnamefont{I.}~\bibnamefont{Sheikin}},
  \bibinfo{author}{\bibfnamefont{D.}~\bibnamefont{Jaccard}},
  \bibinfo{author}{\bibfnamefont{C.}~\bibnamefont{Thessieu}}, \bibnamefont{and}
  \bibinfo{author}{\bibfnamefont{F.}~\bibnamefont{Thomas}},
  \bibinfo{journal}{Phys. Rev. Lett.} \textbf{\bibinfo{volume}{82}},
  \bibinfo{pages}{169} (\bibinfo{year}{1999}).

\bibitem[{\citenamefont{Prohammer and Schachinger}(1987)}]{Prohammer1987}
\bibinfo{author}{\bibfnamefont{M.}~\bibnamefont{Prohammer}} \bibnamefont{and}
  \bibinfo{author}{\bibfnamefont{E.}~\bibnamefont{Schachinger}},
  \bibinfo{journal}{Phys. Rev. B} \textbf{\bibinfo{volume}{36}},
  \bibinfo{pages}{8353} (\bibinfo{year}{1987}).

\bibitem[{\citenamefont{Neto}(2001)}]{Castro2001}
\bibinfo{author}{\bibfnamefont{A.~H.~C.} \bibnamefont{Neto}},
  \bibinfo{journal}{Phys. Rev. Lett.} \textbf{\bibinfo{volume}{86}},
  \bibinfo{pages}{4382} (\bibinfo{year}{2001}).

\bibitem[{\citenamefont{Johannes et~al.}(2006)\citenamefont{Johannes, Mazin,
  and Howells}}]{Johannes2006}
\bibinfo{author}{\bibfnamefont{M.~D.} \bibnamefont{Johannes}},
  \bibinfo{author}{\bibfnamefont{I.~I.} \bibnamefont{Mazin}}, \bibnamefont{and}
  \bibinfo{author}{\bibfnamefont{C.~A.} \bibnamefont{Howells}},
  \bibinfo{journal}{Phys. Rev. B} \textbf{\bibinfo{volume}{73}},
  \bibinfo{pages}{205102} (\bibinfo{year}{2006}).

\bibitem[{\citenamefont{Leroux et~al.}(2012{\natexlab{b}})\citenamefont{Leroux,
  Tacon, Calandra, Cario, M{\'e}asson, Diener, Borrissenko, Bosak, and
  Rodi{\`e}re}}]{Leroux2012}
\bibinfo{author}{\bibfnamefont{M.}~\bibnamefont{Leroux}},
  \bibinfo{author}{\bibfnamefont{M.~L.} \bibnamefont{Tacon}},
  \bibinfo{author}{\bibfnamefont{M.}~\bibnamefont{Calandra}},
  \bibinfo{author}{\bibfnamefont{L.}~\bibnamefont{Cario}},
  \bibinfo{author}{\bibfnamefont{M.~A.} \bibnamefont{M{\'e}asson}},
  \bibinfo{author}{\bibfnamefont{P.}~\bibnamefont{Diener}},
  \bibinfo{author}{\bibfnamefont{E.}~\bibnamefont{Borrissenko}},
  \bibinfo{author}{\bibfnamefont{A.}~\bibnamefont{Bosak}}, \bibnamefont{and}
  \bibinfo{author}{\bibfnamefont{P.}~\bibnamefont{Rodi{\`e}re}},
  \bibinfo{journal}{Phys. Rev. B} \textbf{\bibinfo{volume}{86}},
  \bibinfo{pages}{155125} (\bibinfo{year}{2012}{\natexlab{b}}).

\bibitem[{\citenamefont{Sachdev}(1999)}]{Sachdev1999}
\bibinfo{author}{\bibfnamefont{S.}~\bibnamefont{Sachdev}},
  \emph{\bibinfo{title}{Quantum Phase Transitions}}
  (\bibinfo{publisher}{Cambridge University Press},
  \bibinfo{address}{Cambridge}, \bibinfo{year}{1999}).

\bibitem[{\citenamefont{Coleman and Schofield}(2005)}]{Coleman2005}
\bibinfo{author}{\bibfnamefont{P.}~\bibnamefont{Coleman}} \bibnamefont{and}
  \bibinfo{author}{\bibfnamefont{A.~J.} \bibnamefont{Schofield}},
  \bibinfo{journal}{Nature} \textbf{\bibinfo{volume}{226}},
  \bibinfo{pages}{433} (\bibinfo{year}{2005}).

\bibitem[{\citenamefont{Coleman}(2010)}]{Coleman2010}
\bibinfo{author}{\bibfnamefont{P.}~\bibnamefont{Coleman}},
  \bibinfo{journal}{Phys. Status Solidi B} \textbf{\bibinfo{volume}{247}},
  \bibinfo{pages}{506} (\bibinfo{year}{2010}).

\bibitem[{\citenamefont{Frindt}(1972)}]{Frindt1972}
\bibinfo{author}{\bibfnamefont{R.}~\bibnamefont{Frindt}},
  \bibinfo{journal}{Phys. Rev. Lett.} \textbf{\bibinfo{volume}{28}},
  \bibinfo{pages}{299} (\bibinfo{year}{1972}).

\bibitem[{\citenamefont{Win et~al.}(1991)\citenamefont{Win, Yang, Frindt, and
  Irwin}}]{Qin1991}
\bibinfo{author}{\bibfnamefont{X.}~\bibnamefont{Win}},
  \bibinfo{author}{\bibfnamefont{D.}~\bibnamefont{Yang}},
  \bibinfo{author}{\bibfnamefont{R.}~\bibnamefont{Frindt}}, \bibnamefont{and}
  \bibinfo{author}{\bibfnamefont{J.}~\bibnamefont{Irwin}},
  \bibinfo{journal}{Phys. Rev. B} \textbf{\bibinfo{volume}{44}},
  \bibinfo{pages}{3490} (\bibinfo{year}{1991}).

\bibitem[{\citenamefont{Castellanos-G{\'o}mez
  et~al.}(2010)\citenamefont{Castellanos-G{\'o}mez, Agra{\"i}t, and
  Rubio-Bollinger}}]{Castellanos2010}
\bibinfo{author}{\bibfnamefont{A.}~\bibnamefont{Castellanos-G{\'o}mez}},
  \bibinfo{author}{\bibfnamefont{N.}~\bibnamefont{Agra{\"i}t}},
  \bibnamefont{and}
  \bibinfo{author}{\bibfnamefont{G.}~\bibnamefont{Rubio-Bollinger}},
  \bibinfo{journal}{Appl. Phys. Lett.} \textbf{\bibinfo{volume}{96}},
  \bibinfo{pages}{213116} (\bibinfo{year}{2010}).

\bibitem[{\citenamefont{Galvis et~al.}(2012)\citenamefont{Galvis, Rodi{\`e}re,
  Guillam{\'o}n, Osorio, Rubio-Bollinger, Rodrigo, Cario, Navarro-Moratalla,
  Coronado, S.Vieira et~al.}}]{Augusto2012}
\bibinfo{author}{\bibfnamefont{J.~A.} \bibnamefont{Galvis}},
  \bibinfo{author}{\bibfnamefont{P.}~\bibnamefont{Rodi{\`e}re}},
  \bibinfo{author}{\bibfnamefont{I.}~\bibnamefont{Guillam{\'o}n}},
  \bibinfo{author}{\bibfnamefont{M.}~\bibnamefont{Osorio}},
  \bibinfo{author}{\bibfnamefont{G.}~\bibnamefont{Rubio-Bollinger}},
  \bibinfo{author}{\bibfnamefont{J.}~\bibnamefont{Rodrigo}},
  \bibinfo{author}{\bibfnamefont{L.}~\bibnamefont{Cario}},
  \bibinfo{author}{\bibfnamefont{E.}~\bibnamefont{Navarro-Moratalla}},
  \bibinfo{author}{\bibfnamefont{E.}~\bibnamefont{Coronado}},
  \bibinfo{author}{\bibnamefont{S.Vieira}}, \bibnamefont{et~al.},
  \bibinfo{journal}{ArXiV 1210.2659}  (\bibinfo{year}{2012}).

\bibitem[{\citenamefont{Wang et~al.}(2012)\citenamefont{Wang, Kalantar-Zadeh,
  Kis, Coleman, and Strano}}]{Wang2012}
\bibinfo{author}{\bibfnamefont{Q.~H.} \bibnamefont{Wang}},
  \bibinfo{author}{\bibfnamefont{K.}~\bibnamefont{Kalantar-Zadeh}},
  \bibinfo{author}{\bibfnamefont{A.}~\bibnamefont{Kis}},
  \bibinfo{author}{\bibfnamefont{J.~N.} \bibnamefont{Coleman}},
  \bibnamefont{and} \bibinfo{author}{\bibfnamefont{M.~S.}
  \bibnamefont{Strano}}, \bibinfo{journal}{Nature Nanotechnology}
  \textbf{\bibinfo{volume}{7}}, \bibinfo{pages}{699} (\bibinfo{year}{2012}).

\bibitem[{\citenamefont{Calandra et~al.}(2009)\citenamefont{Calandra, Mazin,
  Mauri}}]{Calandra2009}
\bibinfo{author}{\bibfnamefont{M.} \bibnamefont{Calandra}},
  \bibinfo{author}{\bibfnamefont{I.~I.}~\bibnamefont{Mazin}},
  \bibinfo{author}{\bibfnamefont{F.}~\bibnamefont{Mauri}}, \bibinfo{journal}{Physical Review B} \textbf{\bibinfo{volume}{80}}, \bibinfo{pages}{241108(R)} (\bibinfo{year}{2009}).


\end{thebibliography}

\end{document}